\documentclass[12pt,preprint]{aastex}
\usepackage{color}

\slugcomment{draft of \today}

\shorttitle{Acceleration of CRs at Cosmological Shocks}
\shortauthors{Kang \& Ryu}

\def\ie{{\it i.e.,~}}
\def\kms{~{\rm km~s^{-1}}}
\def\cm3{~{\rm cm^{-3}}}

\def\muG{~{\mu\rm G}}

\begin{document}
\title{DIFFUSIVE SHOCK ACCELERATION AT COSMOLOGICAL SHOCK WAVES}

\author{Hyesung Kang$^1$ and Dongsu Ryu$^2$}
\affil{$^1$Department of Earth Sciences, Pusan National University, Pusan
609-735, Korea: kang@uju.es.pusan.ac.kr\\
$^2$Department of Astronomy and Space Science, Chungnam National
University, Daejeon 305-764, Korea: ryu@canopus.cnu.ac.kr}

\begin{abstract}
We reexamine nonlinear diffusive shock acceleration (DSA) at cosmological shocks 
in the large scale structure of the Universe, incorporating wave-particle interactions that are expected to
operate in collisionless shocks.
Adopting simple phenomenological models for magnetic field amplification (MFA) 
by cosmic-ray (CR) streaming instabilities and Alfv\'enic drift,
we perform kinetic DSA simulations for a wide range of sonic and 
Alfv\'enic Mach numbers
and evaluate the CR injection fraction and acceleration efficiency.
In our DSA model the CR acceleration efficiency is determined mainly by the sonic Mach number $M_s$,
while the MFA factor depends on the Alfv\'enic Mach number and the degree 
of shock modification by CRs.
We show that at strong CR modified shocks, if scattering centers drift with an effective 
Alfv\'en speed in the amplified magnetic field, the CR energy spectrum is steepened
and the acceleration efficiency is reduced significantly, compared to the cases without such effects.
As a result, the postshock CR pressure saturates roughly at $\sim 20$ \% of the shock ram pressure
for strong shocks with $ M_s\ga 10$.  
In the test-particle regime ($M_s\la 3$), it is expected that
the magnetic field is not amplified and the Alfv\'enic drift effects are insignificant,
although relevant plasma physical processes at low Mach number shocks remain largely uncertain.

\end{abstract}


\section{INTRODUCTION}

It is expected that hierarchical gravitational clustering of matter induces 
shock waves in baryonic gas in the large-scale structure (LSS) of the Universe \citep{krj96, miniati00}.
Simulations for the LSS formation suggest that 
strong shocks ($M_s\ga 10$) form in relatively cooler environments 
in voids, filaments, and outside cluster virial radii, 
while weak shock ($M_s \la$ several)
are produced by mergers and flow motions in hotter intracluster media (ICMs) 
\citep{ryuetal03,psej06,kangetal07,skillman08,hoeft08,vazza09,bruggen12}.
Observationally, the existence of such weak shocks in ICMs has been revealed through
temperature jumps in the X-ray emitting gas and Mpc-scale relics with radio spectra
softening downstream of the shock \citep[see][for reviews]{markevitch07,
feretti12}.
These cosmological shocks are the primary means through which the gravitational
energy released during the LSS formation is dissipated into the gas entropy, 
magnetic field, turbulence and nonthermal particles \citep{rkcd08}.

In fact, shocks are ubiquitous in astrophysical environments from the heliosphere to galaxy clusters
and they are thought to be the main `cosmic accelerators' of high energy cosmic-ray (CR) particles \citep{blaeic87}. 
In diffusive shock acceleration (DSA) theory, 
suprathermal particles are scattered by magnetohydrodynamic (MHD) waves and isotropized in the local wave frames,
and gain energy through multiple crossings of the shock \citep{bell78, dru83, maldru01}.
While most postshock thermal particles are advected downstream, 
some suprathermal particles energetic enough to swim against downstream turbulent waves
can cross the shock and be injected into the Fermi first-order process. 
Then these streaming CRs generate resonant waves via two-stream instability 
and nonresonant waves via CR current-driven instability, which in turn amplify
turbulent magnetic fields in the preshock region \citep{bell78,lucek00, bell04,schure12}.
Thin X-ray synchrotron emitting rims observed in several young supernova 
remnants (SNRs) indicate that CR electrons are accelerated to 10-100 TeV and cool radiatively
in the magnetic field of several $100 \muG$ behind
the forward shock \citep[e.g.,][]{parizot06,reynolds12}.
This provides clear evidence for efficient magnetic field amplification during CR acceleration 
at strong CR modified shocks.

These plasma physical processes, \ie injection of suprathermal particles into the CR population, 
excitation of MHD waves and amplification of turbulent magnetic fields via plasma instabilities, 
and further acceleration of CRs via Fermi first-order process are important ingredients of
DSA and should operate
at all types of astrophysical shocks including cosmological shocks in the LSS \citep[e.g.,][]{maldru01, zweibel10,schure12, bruggen12}. 
In addition, relativistic particles can be accelerated stochastically by MHD turbulence,
most likely driven in ICMs of merging clusters \citep{petrosian01,cb05,bl07}.
CRs can be also injected into the intergalactic space by radio galaxies \citep{kdlc01}
and through winds from star-forming galaxies \citep{volk99}, 
and later re-accelerated by turbulence and/or shocks.
Diffuse synchrotron emission from radio halos and relics in galaxy clusters
indicates the presence of GeV electrons gyrating in $\mu$G-level magnetic fields on Mpc scales 
\citep[e.g.,][]{ct02,gf04,wrbh10,krj12}.
On the other hand, non-detection of $\gamma$-ray emission from galaxy clusters by Fermi-LAT 
and VERITAS observations, combined with radio halo observations, puts rather strong constraints 
on the CR proton population and the magnetic field strength in ICMs, if one adopts 
the ``hadronic" model, in which inelastic collisions of CR protons with ICM protons produce 
the radio emitting electrons and
the $\pi^0$ decay \citep{acke10, ddcb10,jeltema11, arlen12}.
Alternatively, in the ``re-acceleration" model, in which those secondary electrons 
produced by p-p collisions are accelerated further by MHD turbulence in ICMs,
the CR proton pressure not exceeding a few \% of the
gas thermal pressure could be consistent with both the Fermi-LAT upper limits from the 
GeV $\gamma$-ray flux and the radio properties of cluster halos \citep{brunetti12}.

Recently, amplification of turbulent magnetic fields via plasma instabilities
and injection of CR protons and electrons at non-relativistic 
collisonless shocks have been studied, using Particle-in-Cell (PIC) and hybrid plasma 
simulations \citep[e.g.][]{riqu09,riqu11,guo10,garat12}. 
In PIC simulations, the Maxwell's equations for electric and magnetic fields are solved 
along with the equations of motion for ions and electrons, so
the full wave-particle interactions can be followed from first principles.
However, extremely wide ranges of length and time scales need to be resolved mainly
because of the large proton to electron mass ratio.
In hybrid simulations, only the ions are treated kinetically
while the electrons are treated as a neutralizing, massless fluid, alleviating
severe computational requirements.
However, it is still prohibitively expensive to simulate the full extent of DSA
from the thermal energies of background plasma to the relativistic energies of cosmic rays, 
following the relevant plasma interactions at the same time. 
So we do not yet have full understandings of injection and diffusive scattering of CRs
and magnetic field amplification (MFA) to make precise quantitative predictions for DSA.
Instead, most of kinetic DSA approaches, in which the diffusion-convection equation
for the phase-space distribution of particles is solved, commonly adopt 
phenomenological models that may emulate some of those processes 
\citep[e.g.,][]{kjg02,bkv09,pzs10,lee12,capri12,kang12}.   
Another approximate method is a steady-state Monte Carlo simulation approach, in which 
parameterized models for particle diffusion, growth of self-generated MHD turbulence,
wave dissipation and plasma heating are implemented \citep[e.g.,][]{vladimirov08}.

In our previous studies, we performed DSA simulations of CR protons at cosmological shocks, 
assuming that the magnetic field strength is uniform in space and constant in time,
and presented the time-asymptotic values
of fractional thermalization, $\delta(M_s)$, and fractional
CR acceleration, $\eta(M_s)$, as a function of the sonic Mach number $M_s$ \citep{kj07, krj09}.
These energy dissipation efficiencies were adopted in a post-processing step 
for structure formation simulations in order to estimate the CR generation at cosmological shocks 
\citep[e.g.,][]{skillman08, vazza09}.
Recently, \citet{vazza12} have used those efficiencies to include self-consistently 
the CR pressure terms in the gasdynamic conservation equations for cosmological simulations. 
In this paper, we revisit the problem of DSA efficiency at cosmological shocks, 
including phenomenological models for MFA
and drift of scattering centers with Alfv\'en speed in the {\it amplified magnetic field}.
Amplification of turbulent magnetic fields driven by CR streaming instabilities is included through an approximate, 
analytic model suggested by \citet{capri12}. 
As in our previous works, a thermal leakage injection model and
a Bohm-like diffusion coefficient ($\kappa(p) \propto p$) are adopted as well.

This paper is organized as follows.
The numerical method and phenomenological models for plasma physical processes in DSA theory,
and the model parameters for cosmological shocks are described in Section 2.  
We then present the detailed simulation results in Section 3 
and summarize the main conclusion in Section 4.

\section{DSA MODEL}

In the diffusion approximation, where the pitch-angle distribution of CRs 
is nearly isotropic,
the Fokker-Plank equation of the particle distribution function is reduced to the following 
diffusion-convection equation: 
\begin{equation}
{\partial f \over \partial t}  + (u+u_w) {\partial f \over \partial x}
= {p \over 3} {{\partial (u+u_w)} \over {\partial x}} 
{{\partial f} \over {\partial p}} 
+ {\partial \over \partial x} \left[\kappa(x,p)
{\partial f \over \partial x} \right],
\label{diffcon}
\end{equation}
where $f(x,p,t)$ is the isotropic part of the pitch-angle averaged CR distribution function,
$\kappa(x,p)$ is the spatial diffusion coefficient
along the direction parallel to the mean magnetic field
and $u_w$ is the drift speed of local Alfv\'enic wave turbulence 
with respect to the plasma \citep{skill75}.
Here, we consider quasi-parallel shocks in one-dimensional planar geometry,
in which the mean magnetic field is roughly parallel to the flow direction.
The flow velocity, $u$, is calculated by solving the momentum
conservation equation with dynamical feedback of the CR pressure and self-generated
magnetic fields,
\begin{equation}
{\partial (\rho u) \over \partial t}  +  {\partial (\rho u^2 + P_g + P_c + P_B) \over \partial x} = 0.
\label{mocon}
\end{equation}
The CR pressure, $P_c$, is calculated self-consistently with the CR distribution function $f$,
while the magnetic pressure, $P_B$, is calculated according to our phenomenological model for
MFA (see Section 2.4)
rather than solving the induction equation \citep{capri09}.
We point that the dynamical effects of magnetic field are not important
with $P_B \la 0.01\rho_0u_s^2$.
The details of our DSA numerical code, 
the CRASH (Cosmic-Ray Amr SHock), can be found in \citet{kjg02}. 

\subsection{Thermal Leakage Injection}

Injection of protons from the postshock thermal pool into the CR population via wave-particle
interactions is expected to depend on several properties of the shock, including the sonic and Alfv\'enic Mach numbers, the obliquity angle of mean magnetic field, and the strength of pre-existing and
self-excited MHD turbulence.
As in our previous studies, we adopt a simple phenomenological model in which particles 
above an
''effective'' injection momentum $p_{\rm inj}$ get
injected to the CR population:
\begin{equation}
p_{\rm inj} \approx 1.17 m_p u_2 \left(1+ {1.07 \over \epsilon_B} \right),
\label{pinj}
\end{equation}
where $\epsilon_B = B_0/B_{\perp}$ is the ratio of
the mean magnetic field along the shock normal, $B_0$, to
the amplitude of the postshock MHD wave turbulence, $B_{\perp}$ \citep{maldru01,kjg02}.
This injection model reflects plasma physical arguments 
that the particle speed must be several times larger than 
the downstream flow speed, $u_2$, depending on the strength of MHD wave turbulence,
in order for suprathermal particles to leak upstream across the shock transition layer.
Since the physical range of the parameter $\epsilon_B$ is not tightly constrained,
we adopt $\epsilon_B =0.25$ as a canonical value, which results in the injected particle
fraction, $\xi=n_{\rm cr,2}/n_2\sim 10^{-4}-10^{-3}$ for $M_s\ga 3$ (see Figure 3 below).
Previous studies showed that DSA saturates for $\xi \ga 10^{-4}$,
so the acceleration efficiency obtained here may represent an upper limit for 
the efficient injection regime \citep[e.g.,][]{kjg02,capri12}.
In fact, this injection fraction is similar to the commonly adopted values for nonlinear
DSA modeling of SNRs \citep[e.g.,][]{bkv09}.
If we adopt a smaller value of $\epsilon_B$ for stronger wave turbulence,
$p_{\rm inj}$ has to be higher, leading to a smaller injection fraction and a lower
acceleration efficiency.

\subsection{Bohm-like Diffusion Model}

In our model, turbulent MHD waves are self-generated efficiently by plasma instabilities driven by
CRs streaming upstream in the shock precursor, so we can assume
that CR particles are resonantly scattered by Alfv\'en waves with fully saturated spectrum.
Then the particle diffusion can be approximated by a Bohm-like diffusion coefficient, 
$\kappa_B \sim (1/3) r_g v$, 
but with flattened non-relativistic momentum dependence \citep{kj07}:
\begin{equation}
\kappa(x,p) = \kappa^* {B_0 \over B_{\parallel}(x)} \cdot {p \over m_pc}, 
\label{Bohm}
\end{equation}
where $\kappa^*= m_p c^3/(3eB_0)= (3.13\times 10^{22} {\rm cm^2s^{-1}}) B_0^{-1}$,
and $B_0$ is the magnetic field strength far upstream expressed in units of $\mu$G.
The strength of the parallel component of local magnetic field, $B_{\parallel}(x)$, will be
described in the next section. 
Hereafter, we use the subscripts `0', `1', and `2' to denote
conditions far upstream of the shock, immediate upstream and downstream of the subshock, respectively.

\subsection{Magnetic Field Amplification}

It was well known that CRs streaming upstream in the shock precursor excite resonant
Alfv\'en waves with a wavelength ($\lambda$) comparable with the CR gyroradius ($r_g$), and 
turbulent magnetic fields can be amplified into the nonlinear regime (\ie $\delta B \gg B_0$) \citep{bell78,lucek00}. 
Later, it was discovered that the nonresonant ($\lambda\ll r_g$), fast-growing instability driven 
by the CR current ($j_{cr}=e n_{cr} u_s$) can amplify the magnetic field by orders of magnitude,
up to the level consistent with the thin X-ray rims at SNRs \citep{bell04}.
Several plasma simulations have shown that both $B_{\parallel}/B_0$
and $B_{\perp}/B_0$ can increase by a factor of up to $\sim 10-45$ via the Bell's CR current-driven instability \citep{riqu09, riqu10,ohira09}.
Moreover, it was suggested that long-wavelength magnetic fluctuations can grow as well
in the presence of short-scale, circularly-polarized Alfv\'en waves excited by the Bell-type instability \citep{bykov11}.
Recently, \citet{roga12} have also shown that large-scale magnetic fluctuations can grow along
the original field by the $\alpha$ effect driven by the nonresonant instability and
both the parallel and perpendicular components can be further amplified.
There are several other instabilities that may amplify the turbulent magnetic field 
on scales greater than the CR gyroradius such as the firehose, filamentation, and acoustic
instabilities \citep[e.g.][]{beresnyak09,dru12,schure12}.
Although Bell's (2004) original study assumed parallel background magnetic field,
it turns out that the non-resonant instability operates for all shocks, regardless of the
inclination angle between the shock normal and the mean background magnetic
field \citep{schure12},  
and so the isoptropization of the amplified magnetic field
can be a reasonable approximation \citep{riqu09,roga12}.
  
Here, we adopt the prescription for MFA due to CR streaming instabilities that was suggested by \citet{capri12}, 
based on the assumption of isotropization of the amplified magnetic field and the effective Alfv\'en speed in the local, amplified field: $\delta B^2/(8\pi \rho_0 u_s^2)=(2/25)(1-U^{5/4})^2 U^{-1.5}$,
where $\bold{\delta B} = \bold{B} - \bold{B_0}$ and $U= (u_s-u)/u_s$ is the flow speed in the shock rest frame 
normalized by the shock speed $u_s$.
In the test-particle regime where the flow structure is not modified,
the upstream magnetic field is not amplified in this model (\ie $U(x)=1$).
In the shock precursor ($x > x_s$, where $x_s$ is the shock position), the MFA factor becomes
\begin{equation}
{\delta B(x)^2 \over B_0^2} = {4\over 25}M_{A,0}^2 { {(1-U(x)^{5/4})^2}\over U(x)^{3/2}},
\label{Bpre}
\end{equation}
where $M_{A,0}= u_s/v_{A,0}$ is the Alfv\'enic Mach number 
for the far upstream Alfv\'en speed, and $v_{A,0}= B_0/ \sqrt{4\pi \rho_0}$.
This model predicts that MFA increases with $M_{A,0}$
and the precursor strength (\ie degree of shock modification by CRs) \citep{vladimirov08}. 
In the case of a ``moderately modified'' shock, in which the immediate preshock speed is $U_1\approx 0.8$, for example, the amplified magnetic pressure increases to 
$\delta B_1^2/8\pi \approx 6.6\times 10^{-3} \rho_0 u_s^2$ 
and the amplification factor scales as $\delta B_1/B_0 \approx 0.12 M_{A,0}$.
We will show in the next section that the shock structure is modified only moderately
owing to the Alfv\'enic drift, so the magnetic field pressure is less than a few \% of the 
shock ram pressure even at strong shocks ($M_s\ga 10$).

For the highest Mach number model considered here, $M_s= 100$, the preshock amplification factor becomes $\delta B_1/B_0 \approx 100$, which is somewhat larger than what was found in the plasma 
simulations for the Bell-type current-driven instability \citep{riqu09,riqu10}.
Considering possible MFA beyond the Bell-type instability by other large-scale instabilities
\citep[e.g.][]{bykov11,roga12,schure12}, this level of MFA may not be out of reach.
Note that this recipe is intended to be a heuristic model that may represent qualitatively the
MFA process in the shock precursor.

Assuming that the two perpendicular components of preshock magnetic fields are completely isotropized
and simply compressed across the subshock, 
the immediate postshock field strength can be estimated by
\begin{equation}
B_2/B_1=\sqrt{1/3+2/3(\rho_2/\rho_1)^2}.
\label{B2}
\end{equation}
We note that the MFA model described in equations (\ref{Bpre})-(\ref{B2}) is also used
for the diffusion coefficient model given by equation (\ref{Bohm}). 

\subsection{Alfv\'enic Drift}

Resonant Alfv\'en waves excited by the cosmic ray streaming are pushed by the
CR pressure gradient ($\partial P_c/\partial x$) and
propagate against the underlying flow in the shock precursor \citep[e.g.][]{skill75,bell78}.
The mean drift speed of scattering centers is commonly approximated as the Alfv\'en speed, \ie
$u_{w,1}(x) \approx + v_A \approx B(x) / \sqrt{4\pi \rho(x)}$, pointing upstream away from the shock, 
where $B(x)$ is the local, amplified magnetic field strength estimated by equation (\ref{Bpre}).
For isotropic magnetic fields, the parallel component would be roughly $B_{\parallel}\approx B(x)/\sqrt{3}$.
But we simply use $B(x)$ for the effective Alfv\'en speed,
since the uncertainty in this model is probably greater than the factor of $\sqrt{3}$
(see Section 3 for a further comment on this factor).
In the postshock region the Alfv\'enic turbulence is probably relatively balanced, 
so the wave drift can be ignored, that is, $u_{w,2} \approx 0$ \citep{jon93}.
Since the Alfv\'enic drift reduces the velocity difference between upstream and 
downstream scattering centers, compared to that of the bulk flow,
the resulting CR spectrum becomes softer than estimated without considering
the wave drift. 
Here, we do not consider loss of turbulent magnetic energy and gas heating 
due to wave dissipation in order to avoid
introducing additional free parameters to the problem. 

\subsection{Set-up for DSA Simulations}

Previous studies have shown that the DSA efficiency depends primarily on the shock sonic Mach number \citep{kangetal07}.
So we considered shocks with a wide range of the sonic Mach number, $M_s=1.5-100$,
propagating into the intergalactic medium (IGM) of different temperature phases, $T_0=10^4-5\times 10^7$ K \citep{krcs05}. Then, the shock speed is given by $u_s=(150\kms)M_s (T_0/10^6{\rm K})^{1/2}$.

We specify the background magnetic field strength by setting the so-called plasma beta, $\beta_P=P_g/P_B$,
the ratio of the gas pressure to the magnetic pressure.
So the upstream magnetic field strength is given as $B_0^2 = 8\pi P_g/\beta_P$,
where $\beta_P \sim 100$ is taken as a canonical value in ICMs \citep[see, e.g.,][]{rkcd08}.
Then, the ratio of the background Alfv\'en speed to the sound speed, $v_{A,0}/c_s=\sqrt{2/(\beta_P \gamma_g)}$ 
(where $\gamma_g$ is the gas adiabatic index), which determines the significance of Alfv\'enic drift,
depends only on the parameter $\beta_P$. 
Moreover, the upstream Alfv\'enic Mach number, $M_{A,0}=u_s/v_{A,0}=M_s\sqrt{\beta_P \gamma_g/2}$, controls the magnetic field
amplification factor as given in equation (\ref{Bpre}).
For $\beta_P=100$ and $\gamma_g=5/3$, the background Aflv\'en speed is about 10 \% of the sound speed,
\ie $v_{A,0}=0.11c_s$ (independent of $M_s$ and $T_0$), and $M_{A,0}=9.1M_s$.
For a higher value $\beta_P$ (\ie weaker magnetic fields), of course, 
the Alfv\'enic drift effect will be less significant.

With a fixed value of $\beta_P$, the upstream magnetic field strength 
can be specified by the upstream gas pressure, $n_{H,0} T_0$, 
as follow:
\begin{equation}
B_0 = 0.28 \muG \left({ {n_{H,0} T_0} \over {10^3\cm3 {\rm K}}}\right)^{1/2}
\left( {100\over \beta_P}\right)^{1/2}.
\label{b0}
\end{equation}
We choose the hydrogen number density, $n_{H,0}=10^{-4} \cm3$,
as the fiducial value to obtain specific 
values of magnetic field strength shown in Figures 1 - 2 below.
But this choice does not affects the time asymptotic results shown 
in Figures 3 - 4,
since the CR modified shock evolves in a self-similar manner and the time-asymptotic states depend primarily on
$M_s$ and $M_{A,0}$, independent of the specific value of $B_0$.

Since the tension in the magnetic field lines hinders Bell's CR current-driven instability,
MFA occurs if the background field strength satisfies the condition, $B_0 < B_s= (0.87 \muG) (n_{cr}u_s)^{1/2}$ \citep{zweibel10}.
For a typical shock speed of $u_s\sim 10^3 \kms$ formed in the IGM with $n_{H,0}\sim 10^{-6}-10^{-4} \cm3$ with the CR injection fraction, $\xi\sim 10^{-4}-10^{-3}$,
the maximum magnetic field for the growth of nonresonant waves is roughly $B_s\sim 0.1-1 \muG$.
The magnetic field strength estimated by equation (\ref{b0}) is $B_0\approx 0.28\muG$ for $n_{H,0}=10^{-4} \cm3$ and $T_0=10^7$ K (ICMs) and
 $B_0\approx 10^{-3} \muG$ for $n_{H,0}=10^{-6} \cm3$ and $T_0=10^4$ K (voids).
Considering the uncertainties in the model and the parameters,
it seems reasonable to assume that MFA via CR streaming instabilities can be effective 
at cosmological shocks in the LSS \citep{zweibel10}.

In the simulations, the diffusion coefficient, $\kappa^*$ in equation (\ref{Bohm}), can be normalized with 
a specific value of $\kappa_o$. 
Then, the related length and time scales are given as $l_o=\kappa_o/u_s$ 
and $t_o=\kappa_o/u_s^2$, respectively. 
Since the flow structure and the CR pressure approach the time-asymptotic self-similar states,
a specific physical value of $\kappa_o$ matters only in the determination
of $p_{\rm max}/m_p c \approx 0.1 u_s^2 t/\kappa^* $ at a given simulation time. 
For example, with $\kappa_o=10^6\kappa^*$, the highest momentum reached at time $t$
becomes $p_{\rm max}/m_p c \approx 10^5(t/t_o)$.

It was suggested that non-linear wave damping and dissipation 
due to ion-neutral collisions may weaken stochastic scatterings, 
leading to slower acceleration and escape of highest energy particles from the shock \citep{pz05}.
Since these processes are not well understood in a quantitative way,
we do not include wave dissipation in the simulations.
Instead we implement a free escape boundary (FEB) at an upstream location
by setting $f(x_{\rm FEB},p)=0$ at $x_{\rm FEB}= 0.5\ l_o$,
which may emulate the escape of the highest energy particles with 
the diffusion length, $\kappa(p)/u_s \ga x_{\rm FEB}$.
Under this FEB condition, the CR spectrum and the shock structure including the precursor
approach the time-asymptotic states in the time scale of $t/t_o\sim 1$ \citep{kang12}.

As noted in the introduction, CR protons can be accelerated by
merger and accretion shocks, injected into the intergalactic space by star forming galaxies 
and active galaxies, and
accelerated by turbulence. Because of long life time and slow diffusion,
CR protons should be accumulated in the LSS over cosmological times.
So it seems natural to assume that ICMs contains pre-existing populations of CR protons.
But their nature is not well constrained, except that
the pressure of CR protons is less than a few \% of the gas thermal
pressure \citep{arlen12,brunetti12}.
For a model spectrum of pre-existing CR protons,
we adopt a simple power-law form,
$f_0(p)=f_{\rm pre}\cdot (p/p_{\rm inj})^{-s}$ for $p \geq p_{\rm inj}$,
with the slope $s = 4.5$, which corresponds to the slope of the test-particle power-law
momentum spectrum accelerated at $M=3$ shocks.
We note that the slope of the CR proton spectrum
inferred from the radio spectral index (\ie $\alpha_R\approx (s-2)/2 $) of
cluster halos ranges $4.5 \la s \la 5$
\citep[e.g.,][]{jeltema11}.
The amplitude, $f_{\rm pre}$, is set by the ratio of the upstream CR to gas
pressure, $R\equiv P_{c,0}/P_{g,0}$, where $R = 0.05$ is chosen as a canonical value.

Table 1 lists the considered models:
the weak shock models with $T_0\ge10^7$ K, 
the strong shock models with $T_0=10^5-10^6$ K,
and the strongest shock models with $T_0=10^4$ K represent shocks formed in hot ICMs, 
in the warm-hot intergalactic medium (WHIM) of filaments, and 
in voids, respectively.
Simulations start with purely gasdynamic shocks initially at rest at $x_s = 0$.

\section{DSA SIMULATION RESULTS}

Figures 1 - 2 show the spatial profiles of magnetic field strength, $B(x)$, and CR pressure,
$P_c(x)$,
and the distribution function of CRs at the shock location, $g_s(p)$, at $t/t_o=0.5, 1, 2$ for models 
without or with pre-existing CRs: from top to bottom panels,
$M_s=3$ and $T_0=5\times 10^7$K, 
$M_s=5$ and $T_0=10^7$K,
$M_s=10$ and $T_0=10^6$K,
and $M_s=100$ and $T_0=10^4$K.
Note that the models with $M_s=3-5$ represent shocks formed in hot ICMs, while those
with $M_s=10$ and 100 reside in filaments and voids, respectively.

The background magnetic field strength corresponds to $B_0=0.63, 0.28, 0.089,$ and $8.9\times 10^{-3}\muG$ for the models with
$M_s=3, 5, 10, $ and 100, respectively, for the fiducial value of $n_{H,0}=10^{-4} \cm3$
(see equation (\ref{b0})).
With our MFA model the postshock field can increase to $B_2 \approx 2-3 \muG$ for all these models,
which is similar to the field strengths observed in radio halos and radio relics.
The postshock CR pressure increases with the sonic Mach number, but saturates at $P_{c,2}/(\rho_0 u_s^2) \approx 0.2$ for $M_s\ga 10$.
One can see that the precursor profile and $g_s(p)$ have reached the time-asymptotic states for $t/t_o\ga 1$ 
for the $M_s=100$ model, while the lower Mach number models are still approaching to steady state at $t/t_o =2$.
This is because in the $M_s=100$ model, by the time $t/t_o\approx 1$ the CR spectrum has extended to 
$p_{\rm max}$ that satisfies the FEB condition.
For strong shocks of $M_s=10-100$, the power-law index, 
$q \equiv -\partial \ln f/\partial \ln p$, is about 4.3 - 4.4 at $p\sim m_p c$ instead of $q=4$,
because the Alfv\'enic drift steepens the CR spectrum.

For the models with pre-existing CRs in Figure 2, the pre-existing population is important
only for weak shocks with $M_s\la 5$, because the injected population dominates in shocks with 
higher sonic Mach numbers.
As mentioned in the Introduction, the signatures of shocks observed in ICMs through X-ray and 
radio observations can be interpreted by low Mach number shocks \citep{markevitch07,feretti12}.
In particular, the presence of pre-existing CRs is expect to be crucial in explaining 
the observations of radio relics \citep{krj12}.

Figure 3 shows time-asymptotic values of downstream gas pressure, $P_{g,2}$, and CR pressure, $P_{c,2}$,
in units of $\rho_0u_s^2$,
density compression ratios, $\sigma_1=\rho_1/\rho_0$ and $\sigma_2=\rho_2/\rho_0$, 
the ratios of amplified magnetic field strengths to background strength, $B_2/B_0$ and $B_1/B_0$, 
and postshock CR number fraction, $\xi=n_{\rm cr,2}/n_2$, as a function of $M_s$
for all the models listed in Table 1. 
We note that for the models without pre-existing CRs (left column) 
two different values of $T_0$ (and so $u_s$) are considered for each of 
$M_s=3,$ 4, 5, 10, 30, and 50 models,
in order to explore the dependence on $T_0$ for a given sonic Mach number. 
The figure demonstrates that the DSA efficiency and the MFA factor are determined
primarily by $M_s$ and $M_{A,0}$, respectively, almost independent of $T_0$.
For instance, the two Mach 10 models with $T_0=10^5$K (open triangle) and $10^6$K (filled triangle)
show the similar results as shown in the left column of Figure 3.
But note that the curves for $P_{cr,2}$ and $\xi$ increase somewhat unevenly near $M_s\approx 4-7$
for the models with pre-existing CRs in the right column,
because of the change in $T_0$ (see Table 1).

At weak shocks with $M_s\la 3$, the injection fraction is $\xi \la 10^{-4}$ and the CR pressure is
$P_{c,2}/\rho_0u_s^2 \la 5\times 10^{-3}$ without pre-existing CRs, while both values
depend on $P_{c,1}$ in the presence of pre-existing CRs.
Since the magnetic field is not amplified in the test-particle regime,
these results remain similar to what we reported earlier in \citet{kr11}.
For a larger value of $\epsilon_B$,
the injection fraction and the CR acceleration efficiency would increase.
As shown in \citet{kj07}, however, $\xi$ and $P_{cr,2}/\rho_0u_s^2$ depend sensitively on 
the injection parameter $\epsilon_B$ for $M_s\la 5$, 
while such dependence becomes weak for $ M_s\ga 10$.
Furthermore, there are large uncertainties in
the thermal leakage injection model especially at weak shocks. 
Thus it is not possible nor meaningful to discuss the quantitative dependence of these
results on $\epsilon_B$, until we obtain more realistic pictures of
the wave-particle interactions through PIC or hybrid plasma
simulations of weak collisionless shocks.

In the limit of large $M_s$, 
the postshock CR pressure saturates at $P_{c,2}\approx 0.2 \rho_0u_s^2$,
the postshock density compression ratio at $\sigma_2 \approx 5$,
and the postshock CR number fraction at $\xi\approx 2\times 10^{-3}$.
The MFA factors are $B_1/B_0\sim 0.12 M_{A,0} \sim M_s$ and $B_2/B_0\sim 3M_s$ for $M_s\ga 5$, 
as expected from equation (\ref{Bpre}).
In \citet{kangetal07} we found that $P_{c,2}\approx 0.55 \rho_0u_s^2$ in the limit of large $M_s$,
when the magnetic field strength was assumed to be uniform in space and constant in time.
Here we argue that MFA and Alfv\'enic drift in the amplified magnetic field steepen the CR spectrum 
and reduce the DSA efficiency drastically.

Again, the presence of pre-existing CRs (right column) enhances the injection fraction
and acceleration efficiency at weak 
shocks of $M_s\la 5$, while it does not affect the results at stronger shocks.
Since the upstream CR pressure is $P_{c,0} = 0.05 P_{g,0}=(0.03/M_s) \rho_0 u_s^2$ in these models, 
the enhancement factor, $P_{c,2}/P_{c,0} \approx 1.5-6$ for $M_s \le 3$.
So the DSA acceleration efficiency exceeds only slightly the adiabatic compression factor,
$\sigma_2^{\gamma_c}$, where $\gamma_c\approx 4/3$ is the adiabatic index of the CR population.

As in \citet{kangetal07}, the gas thermalization and CR acceleration
efficiencies are defined as the ratios of the gas thermal and CR energy fluxes to the shock kinetic energy flux:
\begin{eqnarray}
\delta(M_s) \equiv {{[e_{g,2}-e_{g,0}(\rho_2/\rho_0)^{\gamma_g}]u_2 } \over {(1/2)\rho_0 u_s^3}},~~
\eta(M_s) \equiv {{[e_{c,2}-e_{c,0}(\rho_2/\rho_0)^{\gamma_c}]u_2 } \over {(1/2)\rho_0 u_s^3}},
\label{eff1}
\end{eqnarray}
where $e_{g}$ and $e_{c}$ are the gas thermal and CR energy densities.
The second terms inside the brackets subtract the effect of adiabatic compression occurred at the shock.
Alternatively, the energy dissipation efficiencies not excluding the effect of adiabatic compression 
across the shock can be defined as:
\begin{eqnarray}
\delta^{'}(M_s) \equiv {{[e_{g,2}u_2-e_{g,0}u_0] } \over {(1/2)\rho_0 u_s^3}},~~
\eta^{'}(M_s)   \equiv {{[e_{c,2}u_2-e_{c,0}u_0] } \over {(1/2)\rho_0 u_s^3}},
\label{eff2}
\end{eqnarray}
which may provide more direct measures of the energy generation at the shock.
Note that $\eta=\eta^{'}$ for the models with $P_{c,0}=0$.

Figure 4 shows these dissipation efficiencies for all the models listed in Table 1.
Again, the CR acceleration efficiency saturates at $\eta \approx 0.2$ for $M_s\ga 10$,
which is much lower than what we reported in the previous studies without MFA \citep{ryuetal03, kangetal07}.
The CR acceleration efficiency is $\eta <0.01$ for weak shocks ($M_s\la 3$) if there is
no pre-existing CRs. 
But the efficiency $\eta^{'}$ can be as high as 0.1 even for these weak shocks, 
depending on the amount of pre-existing CRs.
The efficiency $\eta$ for weak shocks is not affected by the new models of MFA and
Alfv\'enic drift,
since the magnetic field is not amplified in the test-particle regime. 

If we choose a smaller value of $\beta_P$, the ratio $v_{A,0}/c_s$ is larger, leading to 
less efficient acceleration due to the stronger Afv\'enic drift effects. 
For example, for $\beta_P\sim 1$ (\ie equipartition
fields), which is relevant for the interstellar medium in galaxies, 
the CR acceleration efficiency in the strong shock limit reduces to $\eta \approx 0.12$  \citep{kang12}.
On the other hand, if we were to choose a smaller wave drift speed, the CR efficiency $\eta$ will increase slightly.
For example, if we choose $u_w \approx 0.3 v_A$ instead of  $u_w \approx v_A$, the value of
$\eta$ in the high Mach number limit would increase to $\sim 0.25$ for the models considered here.

On the other hand, if we choose a smaller injection parameter, for example, $\epsilon_B=0.23$,
the injection fraction reduces from $\xi =2.1\times 10^{-4}$ to $6.2\times10^{-5}$ and 
the postshock CR pressure decreases from  $P_{c,2}/\rho_0 u_s^2=0.076$ to $0.043$ for
the $M_s=5$ model,
while $\xi =2.2\times 10^{-3}$ to $3.3\times10^{-4}$
and $P_{c,2}/\rho_0 u_s^2=0.18$ to $0.14$ for the $M_s=50$ model.
Considering that the CR injection fraction obtained in these simulations ($\xi > 10^{-4}$)
is in the saturation limit of DSA, the CR acceleration efficiency, $\eta$, for $M\ga 10$ 
in Figure 4 should be regarded as an upper limit.

\section{SUMMARY}

We revisited the nonlinear DSA of CR protons at cosmological shocks in the LSS, 
incorporating some phenomenological models for MFA due to CR streaming
instabilities and Alfv\'enic drift in the shock precursor.
Our DSA simulation code, CRASH, adopts the Bohm-like diffusion and 
thermal leakage injection of suprathermal particles into the CR population.

A wide range of preshock temperature, $10^4\le T_0 \le 5\times 10^7$K, is considered
to represent shocks that form in
clusters of galaxies, filaments, and voids.
We found that the DSA efficiency is determined mainly by the sonic Mach number $M_s$,
but almost independent of $T_0$.
We assumed the background intergalactic magnetic field strength, $B_0$, 
that corresponds to the plasma beta $\beta_P=100$. 
This is translated to the ratio of the Alfv\'en speed in the background magnetic field to 
the preshock sound speed, $v_{A,0}/c_s=\sqrt{6/5\beta_P} \approx 0.11$.
Then the Alfv\'enic Mach number $M_{A,0}=\sqrt{5 \beta_P/6}\ M_s$ determines the extent of MFA (\ie $B_1/B_0$),
which in turn controls the significance of Alfv\'enic drift in DSA.
Although the preshock density is set to be $n_{H,0}=10^{-4} \cm3$ just to give a characteristic scale to the magnetic field strength in the IGM,
our results for the CR proton acceleration, such as the dissipation efficiencies, do not depend on
a specific choice of $n_{H,0}$.
If one is interested in CR electrons, which are affected by synchrotron and inverse Compton cooling,
the electron energy spectrum should depend on the field strength $B_0$ and
so on the value of $n_{H,0} T_0$ (see equation (\ref{b0})).

The main results of this study can be summarized as follows:

1) With our phenomenological models for DSA,
the injected fraction of CR particles is $\xi\approx 10^{-4}-10^{-3}$
and the postshock CR pressure becomes $ 10^{-3}\la P_{c,2}/(\rho_0 u_s^2) \la 0.2$
for $3\le M_s\le 100$, if there are no pre-existing CRs.
A population of pre-existing CRs provides seed particles to the Fermi process,
so the injection fraction and acceleration efficiency increase
with the amount of pre-existing CRs at weak shocks.
But the presence of pre-existing CRs does not affect $\xi$ nor $P_{c,2}$ for strong
shocks with $M_s\ga 10$, in which the freshly injected particles dominate over the re-accelerated ones.

2) The nonlinear stage of MFA via
plasma instabilities at collisioness shocks is not fully understood yet. So we adopted a model
for MFA via CR streaming instabilities suggested by \citet{capri12}.
We argue that the CR current, $j_{cr}\sim e \xi \sigma_2 n_{H,0}  u_s$, is high enough to overcome 
the magnetic field tension, so the Bell-type instability can amplify
turbulent magnetic fields at cosmological shocks considered here \citep{zweibel10}.
For shocks with $M\ga 5$, DSA is efficient enough to develop a significant shock precursor
due to the CR feedback,
and the amplified magnetic field strength in the upstream region scales as
$B_1/B_0\approx 0.12 M_{A,0}\approx (\beta_P/100)^{1/2} M_s$.
This MFA model predicts that the postshock magnetic field strength becomes $B_2\approx 2-3 \muG$ for
the shock models considered here (see Table 1).
 
3) This study demonstrates that if scattering centers drift with the effective Alfv\'en speed
in the local, amplified magnetic field, the CR energy spectrum can be steepened 
and the acceleration efficiency is reduced significantly, compared to the cases without MFA.
As a result, the CR acceleration efficiency saturates at $\eta =2e_{c,r}/\rho_0 u_s^3\approx 0.2$
for $M_s \ga 10$,
which is significantly lower than what we reported in our previous study, $\eta \approx 0.55$ 
\citep{kangetal07}. 
We note that the value $\eta$ at the strong shock limit can vary by $\sim 10$ \%,
depending on the model parameters 
such as the injection parameter, plasma beta and wave drift speed. 
Inclusion of wave dissipation (not considered here) will also
affect the extent of MFA and the acceleration efficiency.
This tells us that detailed understandings of plasma physical processes 
are crucial to the study of DSA at astrophysical collisionless shocks.  

4) At weak shocks in the test-particle regime ($M_s\la 3$), 
the CR pressure is not dynamically important enough to generate significant MHD waves,
so the magnetic field is not amplified and the Alfv\'enic drift effects are irrelevant.

5) Finally, we note that the CR injection and the CR streaming instabilities are 
found to be less efficient at quasi-perpendicular shocks \citep[e.g.][]{garat12}.
It is recognized, however, streaming of CRs is facilitated through locally parallel 
inclination of turbulent magnetic fields at the shock surface, so the CR injection can be
effective even at quasi-perpendicular shocks in the presence of
pre-existing large-scale MHD turbulence \citep{giacal05,zank06}.
At oblique shocks the acceleration rate is faster and
the diffusion coefficient is smaller due to drift motion of particles along the shock surface
\citep{jokipii87}.
In fact, the diffusion convection equation (\ref{diffcon}) should be valid for quasi-perpendicular
shocks as long as there exists strong MHD turbulence sufficient enough to keep the pitch 
angle distribution of particles isotropic.
In that case, the time-asymptotic states of the CR shocks should remain the same
even for much smaller $\kappa(x,p)$, as mentioned in Section 2.5.
In addition, the perpendicular current-driven instability is found to be effective
at quasi-perpendicular shocks \citep{riqu10,schure12}.
Thus we expect that the overall conclusions drawn from this study
should be applicable to all non-relativistic shocks, regardless of the magnetic field inclination
angle, although our quantitative estimates for the CR injection and acceleration efficiencies
may not be generalized to oblique shocks with certainty.

\acknowledgements
HK was supported by Basic Science Research Program through
the National Research Foundation of Korea (NRF) funded by the Ministry
of Education, Science and Technology (2012-001065).
DR was supported by the National Research Foundation of Korea
through grant 2007-0093860.
The authors would like to thank D. Capriloi, T. W. Jones, F. Vazza and
the anonymous referee for the constructive suggestions and comments to the paper.
HK also would like to thank Vahe Petrosian and KIPAC for their hospitality during 
the sabbatical leave at Stanford university where a part of the paper
was written.

\clearpage

\begin{deluxetable} {rll}
\tablecaption{Models}
\tablehead{
\colhead {$T_0({\rm K}$)} & \colhead{$P_{c,0}=0$} & \colhead{$P_{c,0}=0.05P_{g,0}$} }

\startdata
$5\times 10^7$ & $M_s=$ 1.5, 2, 3, 4 & $M_s=$ 1.5, 2, 3 \\
$ 10^7$ & $M_s=$ 3, 4, 5 & $M_s=$ 4, 5 \\
$ 10^6$ & $M_s=$ 5, 7, 10 & $M_s=$ 7, 10 \\
$ 10^5$ & $M_s=$ 10, 20, 30, 50 & $M_s=$ 20, 30, 50 \\
$ 10^4$ & $M_s=$ 20, 50, 100 & $M_s=$ 100 \\
\enddata

\end{deluxetable}
\clearpage

\clearpage

\begin{figure}
\vspace{-1cm}
\hskip -1cm
\includegraphics[scale=0.85]{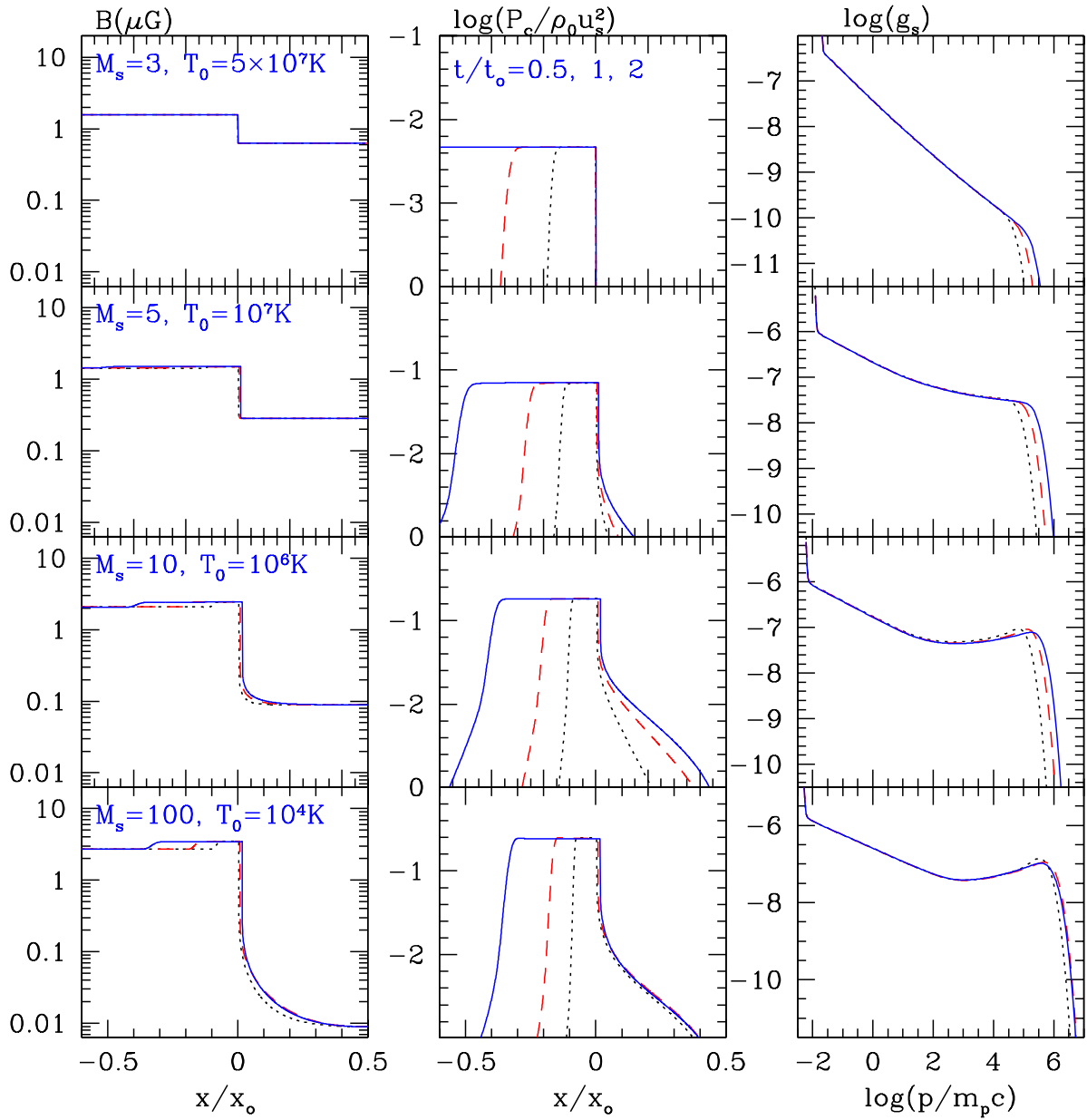}
\vspace{-0.5cm}
\caption{Magnetic field strength, CR pressure profile and the CR distribution at the shock location 
at $t/t_o=0.5$ (dotted lines), 1 (dashed), and 2 (solid) for models without pre-existing CRs. See Table 1 for the model parameters.}
\end{figure}

\clearpage

\begin{figure}
\vspace{-1cm}
\begin{center}
\includegraphics[scale=0.85]{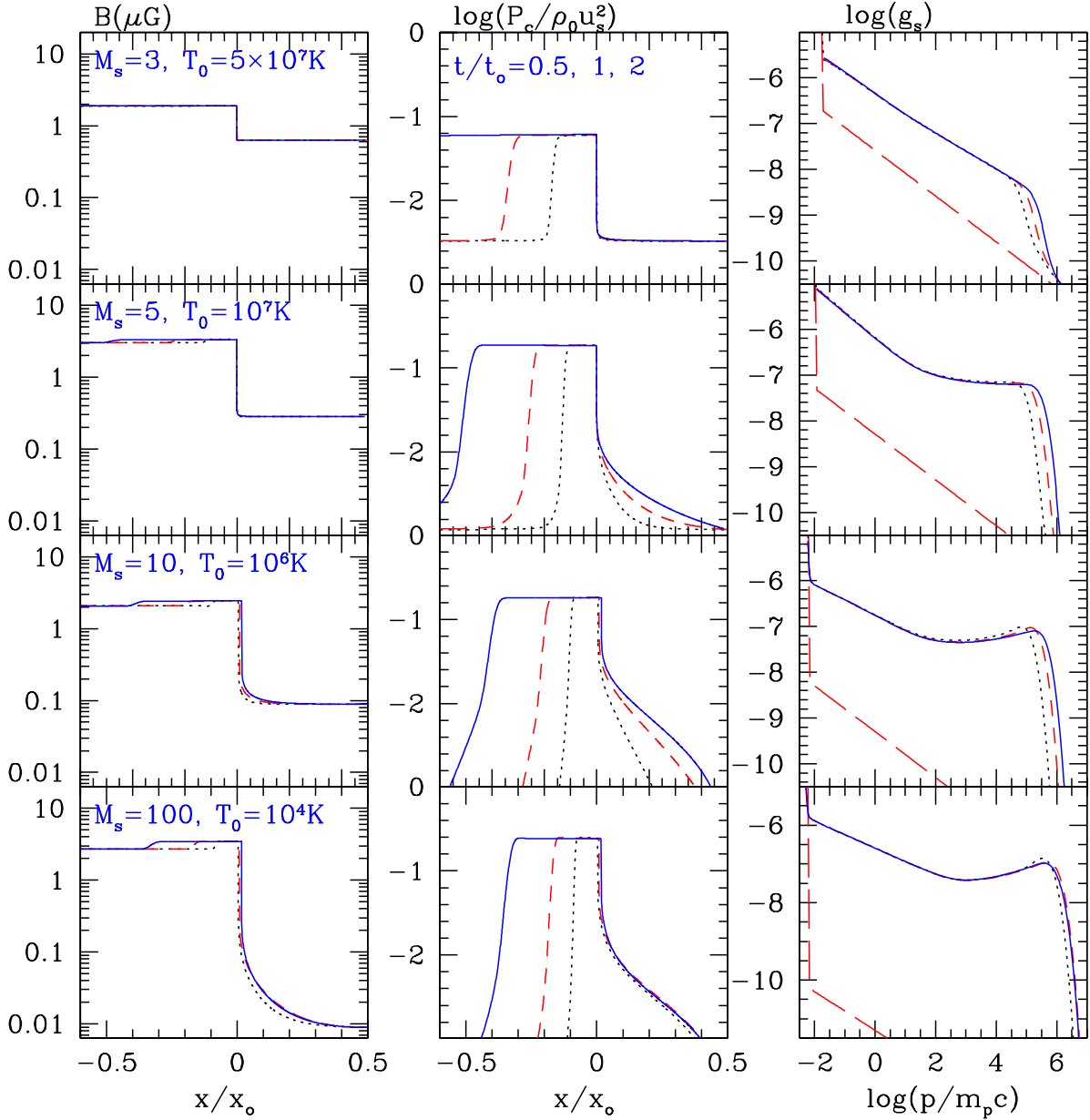}
\end{center}
\vspace{-0.5cm}
\caption{Same as Figure 1 except that models with pre-existing CRs are shown.
The pre-existing population has a power-law spectrum, $f_p\propto p^{-4.5}$
that corresponds to an upstream CR pressure, $P_{c,0}=0.05 P_{g,0}$.
In the right column, the long-dashed lines show the pre-existing population.}
\end{figure}

\clearpage

\begin{figure}
\vspace{-1cm}
\begin{center}
\includegraphics[scale=0.85]{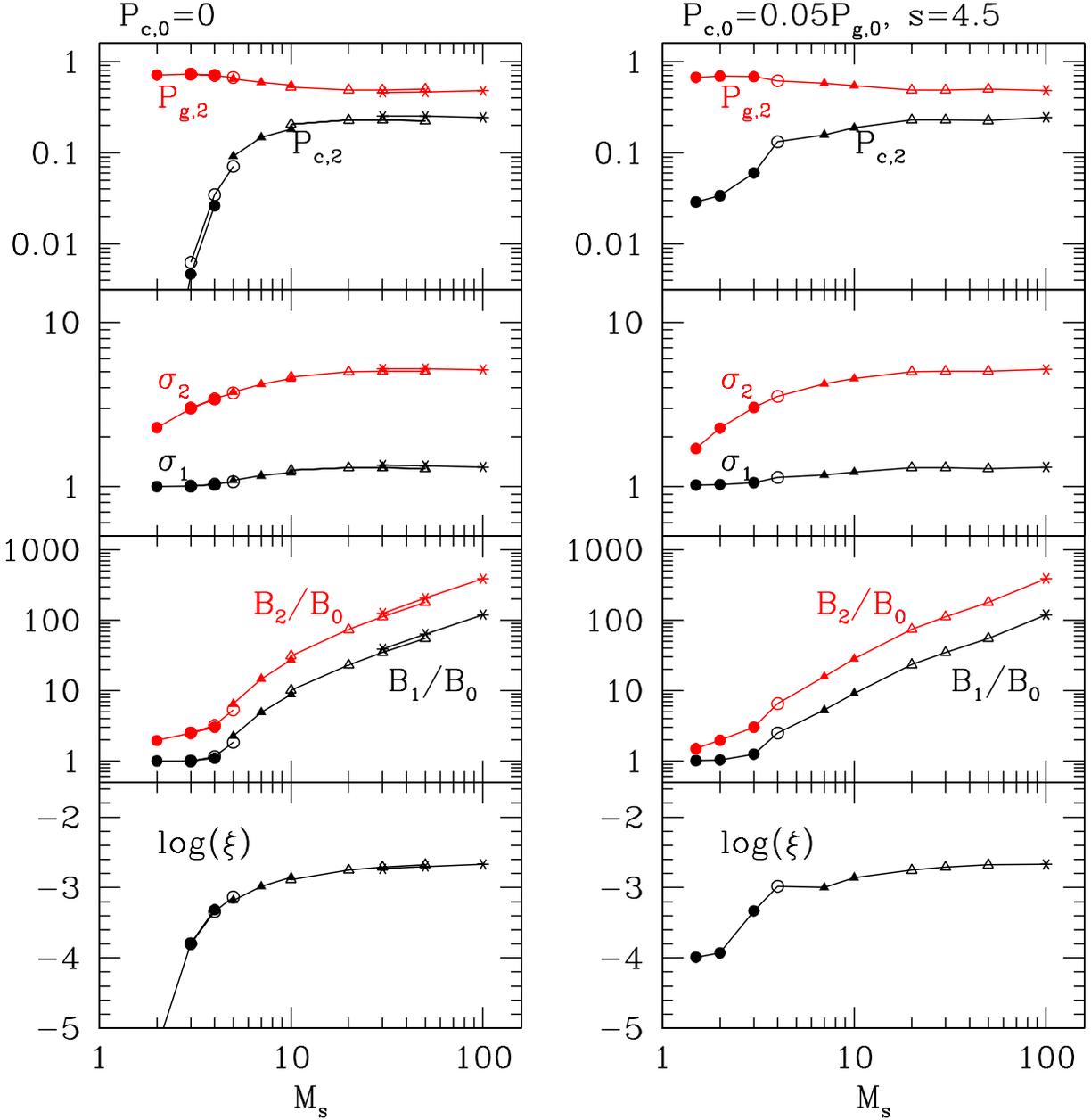}
\end{center}
\vspace{-0.5cm}
\caption{Time-asymptotic values of downstream gas pressure, $P_{g,2}$, and CR pressure, $P_{c,2}$,
in units of $\rho_0u_s^2$,
density compression ratios, $\sigma_1=\rho_1/\rho_0$ and $\sigma_2=\rho_2/\rho_0$, 
the ratios of amplified magnetic field strengths to background strength, $B_2/B_0$ and $B_1/B_0$, 
and postshock CR number fraction, $\xi$, as a function of the sonic Mach number, $M_s$.
The left column shows the cases without pre-existing CRs, while
the right column shows the cases with pre-existing CRs. 
Filled circles are used for the models with $T_0=5\times 10^7$K,
open circles for $T_0=10^7$K, filled triangles for $T_0=10^6$K,
open triangles for $T_0=10^5$K, and stars for $T_0=10^4$K.}
\end{figure}

\clearpage

\begin{figure}
\vspace{-7cm}
\begin{center}
\includegraphics[scale=0.85]{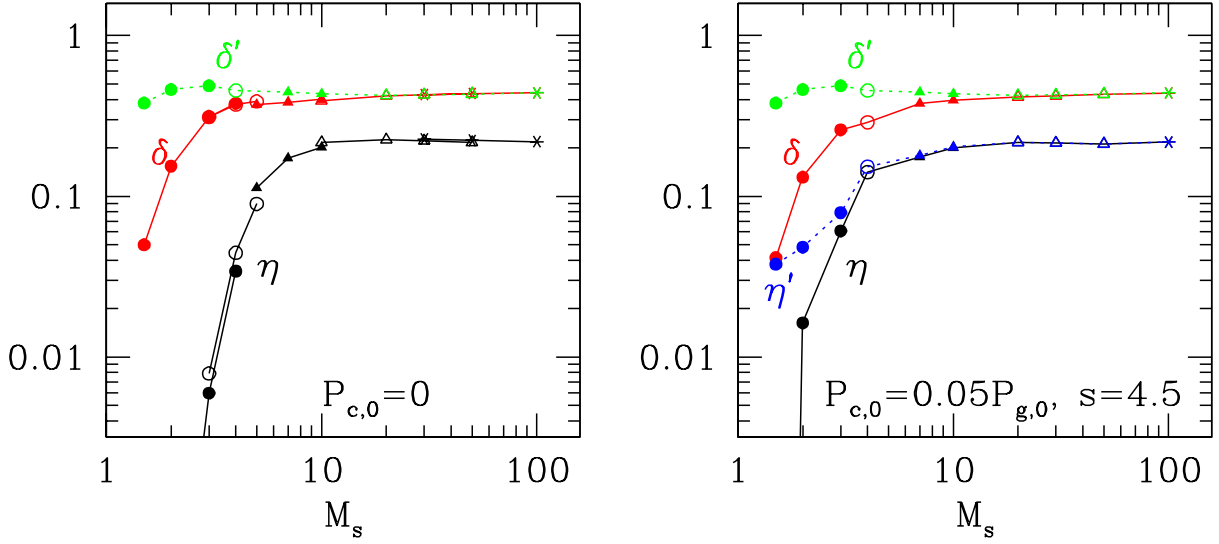}
\end{center}
\vspace{-9cm}
\caption{Shock dissipation efficiencies in the form of gas and CR energies, 
$\delta$ and $\eta$ (solid lines), respectively, in equation (\ref{eff1}) and
$\delta^{'}$ and $\eta^{'}$ (dotted lines) in equation (\ref{eff2}) as a function of the shock sonic Mach number.
For the models with different preshock temperature, $T_0$, 
the same type of symbols are used as in Figure 3.}
\end{figure}

\end{document}